\input{aipcheck}

\documentclass[
    ,final            
  ]
  {aipproc}

\layoutstyle{6x9}

\usepackage{aas_macros}

\def\xmm {\emph{XMM-Newton}}
\def\cxo {\emph{Chandra}}
\def\swift {\emph{Swift}}

\def\src {SGR\,0418+5729}

\def\lum {\mbox{erg s$^{-1}$}}

\begin{document}

\title{SGR\,0418$+$5729: a low-magnetic-field magnetar}

\classification{97.60.Gb -- 98.70.Qy}
\keywords      {pulsars: individual: SGR\,0418$+$5729 -- X-rays: stars -- stars: neutron}

\author{P.~Esposito}{address={INAF--Osservatorio Astronomico di Cagliari, Capoterra (Cagliari), Italy}}
\author{N.~Rea}{address={Institut de Ci\'encies de l'Espai (CSIC--IEEC), Bellaterra (Barcelona), Spain}}
\author{R.~Turolla}{address={Universit\`a di Padova, Dipartimento di Fisica ``Galileo Galilei", Padova, Italy},altaddress={Mullard Space Science Laboratory, UCL, Holmbury St. Mary, Dorking, Surrey, UK}}
\author{G.~L.~Israel}{address={INAF--Osservatorio Astronomico di Roma, Monteporzio Catone (Roma), Italy}}
\author{S.~Zane}{address={Mullard Space Science Laboratory, UCL, Holmbury St. Mary, Dorking, Surrey, UK}}
\author{L.~Stella}{address={INAF--Osservatorio Astronomico di Roma, Monteporzio Catone (Roma), Italy}}
\author{C.~Kouveliotou}{address={NASA/Marshall Space Flight Center, Huntsville, Alabama, USA}}
\author{S.~Mereghetti}{address={INAF--Istituto di Astrofisica Spaziale e Fisica Cosmica Milano, Milano, Italy}}
\author{A.~Tiengo}{address={INAF--Istituto di Astrofisica Spaziale e Fisica Cosmica Milano, Milano, Italy}}
\author{D.~G\"otz}{address={AIM (CEA/DSM-CNRS-UPD), Irfu/SAp, Saclay, Gif-sur-Yvette, France}}
\author{E.~G{\"o}{\u g}{\"u}{\c s}}{address={Sabanc\i\ University, Orhanl\i - Tuzla, \.Istanbul, Turkey}}

\begin{abstract}
Soft gamma-ray repeaters and anomalous X-ray pulsars are a small (but growing) group of X-ray sources characterized by the emission of short bursts and by a large variability in their persistent flux. They are believed to be magnetars, i.e. neutron stars powered by extreme magnetic fields ($10^{14}$--$10^{15}$ G). We found evidence for a magnetar with a low magnetic field, \src, recently detected after it emitted bursts similar to those of soft gamma-ray repeaters. New X-ray observations show that its dipolar magnetic field cannot be greater than $8\times 10^{12}$ G, well in the range of ordinary radio pulsars, implying that a high surface dipolar magnetic field is not necessarily required for magnetar-like activity. The magnetar population may thus include objects with a wider range of magnetic-field strengths, ages and evolutionary stages than observed so far.

\end{abstract}

\maketitle


\section{Introduction}

The surface dipole field strength of a non-accreting pulsar can be estimated equating the rate of rotational kinetic energy loss with the power of the magnetic-dipole radiation. At the neutron-star (magnetic) equator:
\begin{equation}
B  \approx (3Ic^3 \dot{P}P/8\pi^2 R_{\mathrm{NS}}^6)^{1/2} \simeq 3.2\times10^{19}(P \dot{P})^{1/2}~{\rm G}
\label{bspind}
\end{equation}
where $P$ (in s) is the pulsar spin period and $\dot{P}$ (in s s$^{-1}$) its spin-down rate, and we assumed $R_{\mathrm{NS}}=10^6$ cm and $I=10^{45}$ g cm$^2$ for the neutron-star radius and moment of inertia.
The surface dipolar magnetic field strengths inferred in this way are in the range $\sim$$10^{11}$--$10^{13}$ G for most non-recycled pulsars, but they can reach $\sim$$10^{14}$--$10^{15}$ G for a handful of sources which are generally referred to as magnetars.\footnote{The surface dipolar magnetic field of magnetars has been estimated also with several other methods \cite{vietri07,thompson95,thompson01}, all of which give values consistent with those derived from the formula in Eq.~(\ref{bspind}).} To date only a dozen and a half\footnote{For an updated catalogue of SGRs/AXPs see \url{http://www.physics.mcgill.ca/~pulsar/magnetar/main.html}.} of these ultra-magnetized neutron stars are known and, even if the distinction is becoming increasingly blurred, they are generally classified as either soft gamma-ray repeaters (SGRs) or anomalous X-ray pulsars (AXPs). They are all X-ray pulsars with spin periods of 2--12 s, period derivatives of $\sim$$10^{-13}$--$10^{-10}$ s s$^{-1}$ (corresponding to characteristic ages, $\tau_{\rm c} = P/2\dot{P}$, from about 0.2 kyr to 0.2 Myr), and luminosities of $L_{\rm X}\sim 10^{32}$--$10^{36}$ \lum, usually much higher than the rate at which the star loses its rotational energy through spin-down. Magnetars are also characterized by unpredictable outbursts, lasting from days to years, during which they emit characteristic short  bursts of X/$\gamma$-ray photons. Their high luminosities, together with the lack of evidence for accretion from a stellar companion, led to the conclusion that the energy powering the SGR/AXP activity must be stored in their exceptional magnetic fields \cite{thompson95,thompson96}. 

In addition to SGRs and AXPs, two other sources are known to show magnetar-like activity: PSR\,J1846--0258 and PSR\,J1622--4950. The former is a 0.3-s, allegedly rotation-powered, X-ray pulsar, with a magnetic field in the lower end of the magnetar range ($B\sim5\times10^{13}$ G), from which a typical magnetar outburst and short X-ray bursts were detected \cite{gavriil08}. In the latter, flaring radio emission with a flat spectrum (similar to those observed in the two transient radio magnetars; \cite{camilo06,camilo07}) was observed from a 4.3-s radio pulsar with a magnetic field in the magnetar range ($B\sim3\times10^{14}$ G) \cite{levin10}.

In all sources with magnetar-like activity, the dipolar field spans $5\times10^{13}~{\rm G} <B< 2\times10^{15}~{\rm G}$, which is $\sim$10--1000 times the average value in radio pulsars and higher than the quantum electron field, $B_{\rm Q}=m_e^2c^3/e\hbar\simeq 4.4\times10^{13}$ G (at which the electron cyclotron energy equals the rest mass), which is the traditional divide between magnetars and ordinary pulsars. The existence of radio pulsars with $B > B_{\rm Q}$  and showing only normal behavior is an indication that a magnetic field larger than the quantum electron field alone may not be a sufficient condition for the onset of magnetar-like activity \cite{mclaughlin03,kaspi05}. In contrast, so far  the opposite always held: magnetar-like activity was observed only in sources with dipolar magnetic fields stronger than $B_{\rm Q}$.
\vspace{-.4cm}
\section{The source in the focus: \src}

\src\ was discovered on 5 June 2009 when the \emph{Fermi} Gamma-ray Burst Monitor observed two magnetar-like bursts \cite{vanderhorst10}. Follow-up observations with several X-ray satellites showed that \src\ has X-ray pulsations at $\sim$9.1 s \cite{gogus_atel2076_09}, well within the range of periods of magnetar sources, and exhibits all the other peculiarities of magnetars: emission of short X-ray bursts, persistent X-ray luminosity larger than that the rotational energy loss, a spectrum characterized by a thermal plus non-thermal component which softened during the outburst decay, and variable pulse profile \cite{esposito10}.

Several X-ray instruments repeatedly observed \src\ since its discovery and for about 160 days (then, due to solar constraints, the source came out of visibility for the satellites). This campaign allowed an accurate phase-coherent study of the pulsar rotation but no sign of a period derivative was detected \cite{esposito10}. The upper limit on the spin-down rate was $10^{-13}$ s s$^{-1}$ (90\% confidence level), which, according to Eq.~(\ref{bspind}), translates into a limit on the surface dipolar magnetic field of $B<3\times10^{13}$ G \cite{esposito10}. This limit is quite low for a magnetar source, but not abnormally so, it is in fact comparable with the values inferred for the AXP 1E\,1048.0--5937 ($B\sim 6\times 10^{13}$  G) and the magnetar-like pulsar PSR\,J1846--0258.

In July 2010, when \src\ became visible again, we resumed our monitoring campaign by observing the source with \swift, \cxo\ and \xmm\  (see \cite{rea10} for more details). Since pulsations at the known period were clearly detected in all observations, we could use the new X-ray data to extend to a longer baseline the phase-coherent timing solution of \cite{esposito10}. We found that the phase evolution of \src\ is well described by a linear relation $\phi=\phi_0 + 2\pi (t-t_0)/P$, with a best-fitting spin period of 9.07838827(4) s (referred to MJD 54993.0 and to the Solar System barycenter; $\chi_{\rm red}^2 \sim 1.8$ for 18 degrees of freedom). The Fisher test shows that the addition of a quadratic term of the form  $-2\pi \dot P(t-t_0)^2/2 P^2$, which would reflect the presence of a period derivative, is not statistically required. We set an upper limit on the spin-down rate of \src\ of $\dot{P} < 6.0\times10^{-15}$ s s$^{-1}$ (90\% confidence level).
\begin{figure}
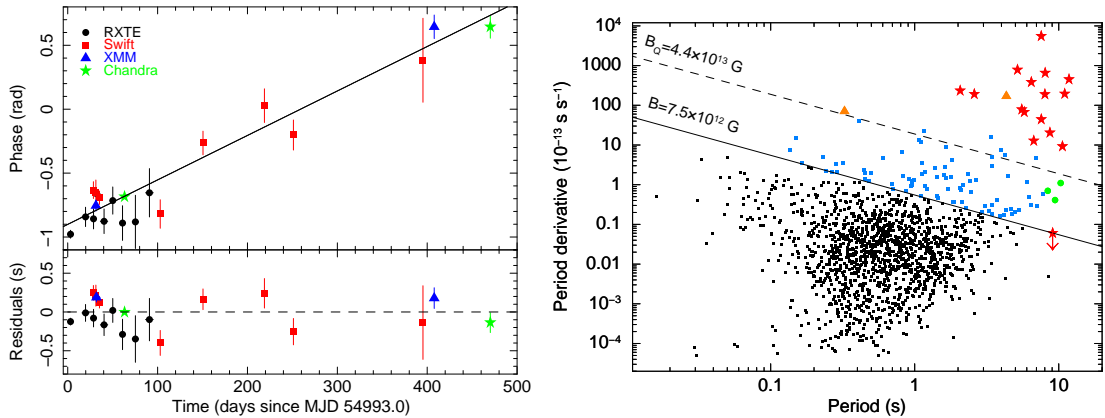

\includegraphics[width=.37\textwidth,angle=-90]{phasefit.eps}
\includegraphics[width=.37\textwidth,angle=-90]{ppdot.eps}
\caption{\emph{Left panel} Rotation phase versus time for the phase-coherent timing solution (a linear function; solid line) for \src\ and (bottom panel) time residuals with respect to the timing solution. \emph{Right panel} $P$--$\dot P$ diagram for all known isolated pulsars (data are from \cite{manchester05}). Black squares represent normal radio pulsars, light-blue squares normal radio pulsars with a magnetic field larger than our limit for \src\ ($7.5\times10^{12}$ G), red stars are the magnetars, orange triangles are the magnetar-like pulsars PSR\,J1846--0258 and PSR\,J1622--4950, and the green circles are the X-ray dim isolated neutron stars (XDINSs; \cite{turolla09}). The solid line marks the 90\% upper limit for the dipolar magnetic field of \src. The value of the electron quantum magnetic field is also reported (dashed line). \label{fig1}
\vspace{-.4cm}}
\end{figure}
\vspace{-.4cm}
\section{Discussion}

The upper limit on the period derivative implies a characteristic age $\tau_c>24$ Myr and a surface dipolar surface magnetic field $B < 7.5\times10^{12}$ G, making \src\ by far the magnetar with the lowest surface dipolar magnetic field yet (much lower also than those of the two magnetar-like pulsars PSR\,J1846--0258 and PSR\,J1622--4950).


Apparently, such a low surface dipolar magnetic field uncomfortably fits in the magnetar model. However, so far we have been considering only the relationship between the surface dipolar magnetic field and magnetar-like activity, but it is likely that the magnetar activity is driven by the magnetic energy stored in the internal toroidal field \cite{thompson95,thompson01}. If the magnetar model as it is currently understood is indeed valid, despite its low surface dipolar field,  \src\ in order to be able to undergo outbursts and emit bursts should harbor a sufficiently-intense internal toroidal component $B_{\rm tor}$. This large internal field can stress the crust and ultimately deforms/cracks the star surface layers, periodically allowing magnetic helicity to be transferred to the external field, thus causing the (repeated) short X-ray bursts and the overall magnetar-like activity \cite{thompson95,tlk02,beloborodov09}. 

The toroidal component of the internal magnetic field cannot be measured directly but can be estimated assuming that the magnetic energy stored in it powers the quiescent emission of \src\ during its entire lifetime, $B_{\rm tor}^2 \approx 6 L_{\rm X} \tau_{\rm c} / R_{\rm NS}^3$ \cite{thompson95}. Assuming a distance to the source of 2 kpc \cite{vanderhorst10,esposito10}, and that the luminosity $L_{\rm X}\sim6.2\times10^{31}$ \lum\ measured with \cxo\ in July 2010 corresponds to the quiescent luminosity, we obtain, for a neutron-star radius of $R_{\rm NS}= 10^6$ cm and a characteristic age $\tau_{\rm c}=24$ Myr,  $B_{\rm tor} \approx 5\times10^{14}$ G. A value of the same order is obtained if the ratio of the toroidal to poloidal field strength is $\sim$50, as in the magneto-thermal evolution scenario \cite{pons09}. In this picture, \src\ may possess a high-enough internal magnetic field to overcome the crustal yield strength (the minimum value to do this should be in excess of $\approx$$10^{14}$ G \cite{thompson95}) and give rise to magnetar-like activity despite its low surface dipolar magnetic field.  However, would the actual surface dipolar magnetic field of \src\ turn out to be much smaller than the present upper limit, this may require to rethink some of the ingredients at the basis of the magnetar scenario.

Finally, we note that a large fraction of the radio pulsar population may have, similarly to \src, magnetar-like internal fields not reflected in their normal dipolar component. Thus, magnetar-like activity may occur in pulsars with a very wide range of magnetic fields and it may fill a continuum in the $P$--$\dot{P}$ diagram (Fig.~\ref{fig1}). Among the presumably rotation-powered pulsars, $\sim$18\% have a dipolar magnetic field higher than the upper limit we derived for \src\ (Fig.~\ref{fig1}). The discovery  of PSR\,J1622--4950 \cite{levin10} on the other hand, suggests that magnetar-like behavior may manifest itself mostly in the radio band. In this framework, our result indicates that a large number of apparently normal pulsars might turn on as magnetars at anytime, regardless of having a surface dipole magnetic field above the quantum limit or not. 
\vspace{-.4cm}
\bibliographystyle{aipproc}   

\begin{thebibliography}{20}
\expandafter\ifx\csname natexlab\endcsname\relax\def\natexlab#1{#1}\fi
\providecommand{\enquote}[1]{``#1''}
\expandafter\ifx\csname url\endcsname\relax
  \def\url#1{\texttt{#1}}\fi
\expandafter\ifx\csname urlprefix\endcsname\relax\def\urlprefix{URL }\fi
\providecommand{\eprint}[2][]{\url{#2}}

\bibitem[{Vietri} et~al.(2007)]{vietri07}
M.~{Vietri}, L.~{Stella}, and G.~L. {Israel}, \emph{\apj} \textbf{661},
  1089 (2007).

\bibitem[{Thompson} and {Duncan}(1995)]{thompson95}
C.~{Thompson}, and R.~C. {Duncan}, \emph{\mnras} \textbf{275}, 255 (1995).

\bibitem[{Thompson} and {Duncan}(2001)]{thompson01}
C.~{Thompson}, and R.~C. {Duncan}, \emph{\apj} \textbf{561}, 980 (2001).

\bibitem[{Thompson} and {Duncan}(1996)]{thompson96}
C.~{Thompson}, and R.~C. {Duncan}, \emph{\apj} \textbf{473}, 322 (1996).

\bibitem[{Gavriil} et~al.(2008)]{gavriil08}
F.~P. {Gavriil}, M.~E. {Gonzalez}, E.~V. {Gotthelf}, et~al., \emph{Science} \textbf{319}, 1802
  (2008).

\bibitem[{Camilo} et~al.(2006)]{camilo06}
F.~{Camilo}, S.~M. {Ransom}, J.~P. {Halpern}, et~al., \emph{\nat} \textbf{442}, 892
  (2006).

\bibitem[{Camilo} et~al.(2007)]{camilo07}
F.~{Camilo}, S.~M. {Ransom}, J.~P. {Halpern}, and J.~{Reynolds}, \emph{\apjl}
  \textbf{666}, L93 (2007).

\bibitem[{Levin} et~al.(2010)]{levin10}
L.~{Levin}, M.~{Bailes}, S.~{Bates}, et~al.,
  \emph{\apjl} \textbf{721}, L33 (2010).

\bibitem[{McLaughlin} et~al.(2003)]{mclaughlin03}
M.~A. {McLaughlin}, I.~H. {Stairs}, V.~M. {Kaspi}, et~al., \emph{\apjl} \textbf{591},
  L135 (2003).

\bibitem[{Kaspi} and {McLaughlin}(2005)]{kaspi05}
V.~M. {Kaspi}, and M.~A. {McLaughlin}, \emph{\apjl} \textbf{618}, L41
  (2005).

\bibitem[{van der Horst} et~al.(2010)]{vanderhorst10}
A.~J. {van der Horst}, V.~{Connaughton}, C.~{Kouveliotou} et~al., \emph{\apjl}
  \textbf{711}, L1 (2010).

\bibitem[{G{\"o}{\u g}{\"u}{\c s}} et~al.(2009)]{gogus_atel2076_09}
E.~{G{\"o}{\u g}{\"u}{\c s}}, P.~{Woods}, and C.~{Kouveliotou}, \emph{Astron.
  Tel.} \textbf{2076} (2009).

\bibitem[{Esposito} et~al.(2010)]{esposito10}
P.~{Esposito}, G.~L. {Israel}, R.~{Turolla}, et~al.,
  \emph{\mnras} \textbf{405}, 1787 (2010).

\bibitem[{Rea} et~al.(2010)]{rea10}
N.~{Rea}, P.~{Esposito}, R.~{Turolla}, et~al., \emph{Science} \textbf{330}, 944 (2010).

\bibitem[{Manchester} et~al.(2005)]{manchester05}
R.~N. {Manchester}, G.~B. {Hobbs}, A.~{Teoh}, and M.~{Hobbs}, \emph{\aj}
  \textbf{129}, 1993 (2005).

\bibitem[{Turolla}(2009)]{turolla09}
R.~{Turolla}, in \emph{Neutron stars and pulsars},
  \emph{ASSL} vol. \textbf{357}, ed. {W.~Becker},
  p. 141 (Springer, 2009).

\bibitem[{Thompson} et~al.(2002)]{tlk02}
C.~{Thompson}, M.~{Lyutikov}, and S.~R. {Kulkarni}, \emph{\apj} \textbf{574},
  332 (2002).

\bibitem[{Beloborodov}(2009)]{beloborodov09}
A.~M. {Beloborodov}, \emph{\apj} \textbf{703}, 1044 (2009).

\bibitem[{Pons} et~al.(2009)]{pons09}
J.~A. {Pons}, J.~A. {Miralles}, and U.~{Geppert}, \emph{\aap} \textbf{496},
  207 (2009).

\end{thebibliography}

\end{document}